\documentclass[aps,twocolumn,preprintnumbers,showpacs,showkeys,nofootinbib%
]{revtex4}
\usepackage{epsfig}
\usepackage{graphicx}
\usepackage{amssymb,amsmath,amsfonts,amsthm,graphicx,psfrag}
\graphicspath{{./Figures/}}                 
\newcommand{\bea}{\begin{eqnarray}}
\newcommand{\eea}{\end{eqnarray}}

\newcommand{\Tab}[1]{Table~\ref{#1}}

\newcommand{\tr}{\operatorname{Tr}}

\begin{document}

\title{ Abelian monopoles in finite temperature lattice $SU(2)$ gluodynamics: first study with improved action}

\author{V.~G.~Bornyakov}
\affiliation{High Energy Physics Institute, 142280 Protvino, Russia \\
and Institute of Theoretical and Experimental Physics, 117259 Moscow, Russia}

\author{A.~G.~Kononenko}
\affiliation{Joint Institute for Nuclear Research, 141980, Dubna, Russia}
\affiliation{Moscow State University, Physics Department, Moscow, Russia}

\begin{abstract}

The properties of the thermal Abelian color-magnetic monopoles in the maximally Abelian gauge are studied in the deconfinement phase of the lattice $SU(2)$ gluodynamics.
To check  universality of the monopole properties we employ the tadpole improved Symanzik action.
The simulated annealing algorithm combined with multiple gauge copies is applied for fixing the maximally Abelian gauge to avoid effects of Gribov copies.
We compute the density, interaction parameters, thermal mass and chemical potential of the thermal Abelian monopoles in the temperature range between $T_c$ and $3T_c$.
In comparison with earlier findings our results for these quantities are improved either with respect to effects of Gribov copies or with respect to lattice artifacts.

\end{abstract}



\maketitle

\section{Introduction}

The signatures of the strong interactions in the quark-gluon matter were found in heavy ion collisions experiments~\cite{Adams:2005dq}.
There are proposals~\cite{Liao:2006ry,Chernodub:2006gu} suggesting that color-magnetic monopoles contribution can explain this rather unexpected property.
These proposals inspired a number of publications devoted to the properties and possible roles of the monopoles in the quark-gluon phase~\cite{Shuryak:2008eq,Ratti:2008jz,D'Alessandro:2007su,Liao:2008jg,D'Alessandro:2010xg,Bornyakov:2011th,Bornyakov:2011eq}.

Lattice gauge theory suggests a direct way to study fluctuations contributing to the Euclidean space functional integral, in particular, the color-magnetic monopoles can be studied.
In a number of papers the evidence was found that the nonperturbative properties of the nonabelian gauge theories such as confinement, deconfining transition, chiral symmetry breaking, etc. are closely related to the Abelian monopoles defined in the maximally Abelian gauge $(MAG)$~\cite{'tHooft:1981ht,Kronfeld:1987ri}.
This was called a monopole dominance~\cite{Shiba:1994ab}.
The drawback of this approach to the monopole studies is that the definition is based on the choice of Abelian gauge.
There are various arguments supporting the statement that the Abelian monopoles found in the MAG are important physical fluctuations surviving the cutoff removal: scaling of the monopole density at $T=0$ according to dimension $3$ for infrared $(percolating)$ cluster~\cite{Bornyakov:2005iy};
Abelian and monopole dominance for a number of infrared physics observables (string tension~\cite{Shiba:1994ab,Suzuki:1989gp,Bornyakov:2005iy}, chiral condensate~\cite{Woloshyn:1994rv}, hadron spectrum~{\cite{Kitahara:1998sj});
monopoles in the MAG are correlated with gauge invariant objects - instantons and calorons~\cite{Ilgenfritz:2006ju,Hart:1995wk}.
It has been recently argued that the MAG is a proper Abelian gauge to find gauge invariant monopoles since t'Hooft-Polyakov monopoles can be identified in this gauge by the Abelian flux, but this is not possible in other Abelian gauges~\cite{Bonati:2010tz}.
Most of these results were obtained for $SU(2)$ gluodynamics but then confirmed for $SU(3)$ theory and QCD~\cite{Arasaki:1996sm,Bornyakov:2003vx}.
Listed above properties of Abelian monopoles survive the continuum limit and removal of the Gribov copy effects.
It is worth noticing that removal of Gribov copy effects changes numerical values of monopole characteristics quite substantially~\cite{Bali:1996dm}.

In this paper we are studying thermal monopoles.
It was shown in Ref.~\cite{Chernodub:2006gu} that thermal monopoles in Minkowski space are associated with Euclidean monopole trajectories wrapped around the temperature direction of the Euclidean volume.
So the density of the monopoles in the Minkowski space is given by the average of the absolute value of the monopole wrapping number.

In~\cite{Liao:2006ry} another approach to study thermal monopole properties in the quark-gluon plasma phase based on the molecular dynamics algorithm was suggested and implemented.
The results for parameters of inter-monopole interaction were found in agreement with lattice results~\cite{Liao:2008jg}.

First numerical investigations of the wrapping monopole trajectories were performed in $SU(2)$ Yang-Mills theory at high temperatures in Refs.~\cite{Bornyakov:1991se} and~\cite{Ejiri:1995gd}.~A more systematic study of the thermal monopoles was performed in Ref.~\cite{D'Alessandro:2007su}.
It was found in~\cite{D'Alessandro:2007su} that the density of monopoles is independent of the lattice spacing, as it should be for a physical quantity.
The density--density spatial correlation functions were computed in~\cite{D'Alessandro:2007su}.
It was shown that there is a repulsive (attractive) interaction for a monopole--monopole (monopole--antimonopole) pairs, which at large distances might be described by a screened Coulomb potential with a screening length of the order of $0.1$ fm.
In Ref.~\cite{Liao:2008jg} it was proposed to associate the respective coupling constant with a magnetic coupling $\alpha_{m}$.
In the paper~\cite{D'Alessandro:2010xg} trajectories which wrap more than one time around the time direction were investigated.
It was shown that these trajectories contribute significantly to a total monopole density at $T$ slightly above $T_c$.
It was also demonstrated that Bose condensation of thermal monopoles, indicated by vanishing of the monopole chemical potential, happens at temperature very close to $T_c$.
However, the relaxation algorithm applied in~\cite{D'Alessandro:2007su} to fix the MAG is a source of the systematic errors due to effects of Gribov copies.
It is known since long ago that these effects are strong in the MAG and results for gauge noninvariant observables can be substantially corrupted by inadequate gauge fixing~\cite{Bali:1996dm}.
For the density of magnetic currents at zero temperature it might be as high as $20\%$.

For nonzero temperature  the effects of Gribov copies were not investigated until recently.
In a recent paper~\cite{Bornyakov:2011eq} this gap was partially  closed.
It was shown that indeed gauge fixing with SA algorithm and $10$ gauge copies per configuration gives rise to the density of the thermal monopoles $20$ to $30\%$ lower (depending on the temperature) than values found in~\cite{D'Alessandro:2007su}.
Large systematic effects due to effects of Gribov copies found in Ref.~\cite{Bornyakov:2011eq} imply that results obtained in earlier papers~\cite{D'Alessandro:2007su,Liao:2008jg,D'Alessandro:2010xg} for the density and other monopole properties can not be considered as
quantitatively precise and need  further independent verification .
The quantitatively precise determination of  such parameters as monopole density, monopole coupling and others is necessary, in particular,
 to verify the conjecture ~\cite{Liao:2006ry} that the magnetic monopoles are weakly interacting (in comparison with electrically charged fluctuations) just above transition but become strongly interacting at high temperatures.
In this paper we use the same gauge fixing procedure as in Refs.~\cite{Bornyakov:2011th,Bornyakov:2011eq} to avoid systematic effects due to Gribov copies.

The careful  study of the finite volume and finite lattice spacing effects was made in \cite{D'Alessandro:2007su}.
We fix our spatial lattice size to $L_s=48$ which was shown in Ref.~\cite{D'Alessandro:2007su} to be large enough to avoid finite volume effects.
We check finite lattice spacing effects comparing results obtained on lattices
with $N_t=4$ and 6 at two temperatures.
Let us emphasize that our studies are computationally much more demanding in comparison with studies undertaken in Refs.~\cite{D'Alessandro:2007su,D'Alessandro:2010xg}, since we produce $10$ Gribov copies per configuration to avoid Gribov copies effect.
For this reason our check of the continuum limit is not as extensive as it was in Refs.~\cite{D'Alessandro:2007su,D'Alessandro:2010xg}.

The important  contribution of this work to the thermal monopole studies is a check of universality.
In studies of magnetic currents at zero temperature it was found~\cite{Bornyakov:2005iy} that the density of the infrared magnetic currents is different for different lattice actions with difference as large as $30\%$.
The conclusion was made that the ultraviolet fluctuations contribute to the infrared density and this contribution has to be removed.
Partial removal was made by the use of the improved action.
In present paper we use the improved lattice action - tadpole improved Symanzik action and compare our results for the density and other quantities with
results obtained with the Wilson action~\cite{D'Alessandro:2007su,Liao:2008jg,D'Alessandro:2010xg,Bornyakov:2011eq}.
We find that the universality holds for monopoles which do not form short range (ultraviolet) dipoles.

We also want to point out that in this work we use more natural procedure of computing monopole correlators in comparison with papers~\cite{D'Alessandro:2010xg}
and~\cite{Bornyakov:2011eq}.
It was mentioned in Ref.~\cite{D'Alessandro:2010xg} that monopole trajectories had a lot of small loops attached to them which were UV noise.
Presence of such loops do not allow to determined monopole spatial coordinates unambiguously for all time slices.
This problem was bypassed in~\cite{D'Alessandro:2010xg,Bornyakov:2011eq} by using only one time slice.
We remove the small loops attached to thermal monopole trajectories and thus we are able to determine the monopole coordinates in every time slice unambiguously. Then we use all time slices to compute the correlators  what allows us to decrease the statistical errors substantially.

\section{Simulation details}

We studied the $SU(2)$ lattice gauge theory with the tadpole improved Symanzik action:

\begin{equation}
S = \beta_{impr}\sum_{pl}S_{pl}-\frac{\beta_{impr}}{20u_{0}^{2}}\sum_{rt}S_{rt}
\label{eq:Symanzik}
\end{equation}

\noindent where $S_{pl}$ and $S_{rt}$ denote plaquette and 1$\times$2 rectangular loop terms in the action:

\begin{equation}
S_{pl,rt} = \frac{1}{2}Tr(1 - U_{pl,rt}),
\label{eq:action_plaquette}
\end{equation}

\noindent  parameter $u_{0}$ is the input tadpole improvement factor taken here equal to the fourth root of the average plaquette P = $\langle\frac{1}{2}U_{pl}\rangle$.
We use the same  code to generate configurations of the lattice gauge field as was used in Ref.~\cite{Bornyakov:2005iy}.
Our calculations were performed on the asymmetric lattices with lattice volume $V=L_t L_s^3$, where $L_{t,s}$ is the number of sites in the time (space) direction.
The temperature $T$ is given by:

\begin{equation}
T = \frac{1}{aL_t}~,
\label{eq:T,Lt,a}
\end{equation}

\noindent where $a$ is the lattice spacing.
To determine the values of $u_{0}$ we used results of Ref.~\cite{Bornyakov:2005iy} either directly or to make interpolation to necessary values of $\beta$.
The critical value of the coupling constant for $L_t=6$ is $\beta_c=3.248$ \cite{Bornyakov:2007}.
For $L_t=6$ the ratio $T/T_c$ was obtained using data for the string tension from Ref.~\cite{Bornyakov:2005iy} again either directly or via interpolation.
For $L_t=4$ ratio $T/T_c$ was taken to be equal to the ratio for $L_t=6$ multiplied by factor $1.5$.
In~\Tab{tab:lattice_size} we provide the information about the gauge field ensembles and parameters used in our study.

The MAG is fixed by finding an extremum of the gauge functional:
\begin{equation}
F_U(g) = ~\frac{1}{4V}\sum_{x\mu}~\frac{1}{2}~\tr~\biggl( U^{g}_{x\mu}\sigma_3 U^{g\dagger}_{x\mu}\sigma_3 \biggr) \;,
\label{eq:gaugefunctional}
\end{equation}
\noindent with respect to gauge transformations $g_x$ of the link variables $U_{x\mu}$:

\begin{equation}
U_{x\mu} \stackrel{g}{\mapsto} U_{x\mu}^{g}
= g_x^{\dagger} U_{x\mu} g_{x+\mu} \; ;
\qquad g_x \in SU(2) \,.
\label{eq:gaugetrafo}
\end{equation}

We apply the simulated annealing (SA) algorithm which proved to be very efficient for this gauge~\cite{Bali:1996dm} as well as for other gauges such as center gauges~\cite{Bornyakov:2000ig} and Landau gauge~\cite{Bogolubsky:2007pq}.
To further decrease the Gribov copy effects we generated $10$ Gribov copies per configuration starting every time gauge fixing procedure from a randomly selected gauge copy of the original Monte Carlo  configuration.

\begin{table}[ht]
\begin{center}
\begin{tabular}{|c|c|c|c|c|c|}
\hline
$\beta$& $u_{0}$ & $L_t$ & $L_s$ & $T/T_c$ & $N_{meas}$ \\
\hline
\hline
3.640  &0.92172  &  4    & 48  & 3.00                  & 300    \\
3.544  &0.91877  &  4    & 48  & 2.50                  & 200    \\
3.480  &0.91681  &  4    & 48  & 2.26                  & 200    \\
3.410  &0.91438  &  4    & 48  & 2.00                  & 200    \\
3.248  &0.90803  &  4    & 48  & 1.50                  & 254    \\
\hline
\hline
3.640  &0.92172  &  6    & 48  & 2.00                  & 200    \\
3.480  &0.91681  &  6    & 48  & 1.50                  & 200    \\
3.400  &0.91402  &  6    & 48  & 1.31                  & 270    \\
3.340  &0.91176  &  6    & 48  & 1.20                  & 200    \\
3.300  &0.91015  &  6    & 48  & 1.10                  & 203    \\
3.285  &0.90954  &  6    & 48  & 1.07                  & 206    \\
3.265  &0.90867  &  6    & 48  & 1.03                  & 200    \\
\hline

\end{tabular}
\end{center}
\caption{
Values of $\beta$, parameter $u_0$, lattice sizes, temperature, number of measurements used in this paper.
}
\label{tab:lattice_size}
\medskip
\noindent
\end{table}

\section{Monopole Density}

The monopole current is defined on the links $\{ x, \mu \}^*$ of the dual lattice and take integer values $j_{\mu}(x)=0, \pm 1, \pm 2$.
The monopole currents form closed loops combined into clusters.
Wrapped clusters are closed through the lattice boundary.
The wrapping number $N_{wr} \in Z$ of a given cluster is defined by:

\begin{equation}
N_{wr} = \frac{1}{L_t} \sum_{j_4(x) \in cluster} j_4(x)\,
\label{eq:wrapping_number}
\end{equation}

The density $\rho$ of the thermal monopoles is defined as follows:

\begin{equation}
\rho = \frac{\langle~ \sum_{clusters}|N_{wr}| ~\rangle }{L_s ^3 a^3}\,
\label{eq:lattice_density}
\end{equation}

\noindent where the sum is taken over all wrapped clusters for a given configuration.

Following Ref.~\cite{Chernodub:2006gu} we distinguish two regions of temperatures: low temperature region $T\lesssim2T_{c}$ and high temperatures $T\gtrsim2T_{c}$.
In Ref.~\cite{Chernodub:2006gu} it was proposed that at low temperatures the density of monopoles is almost insensitive to
temperature thus indicating that the monopoles form a dense liquid while at high temperature the monopole density has a power dependence
on temperature.
These statements were based on the results obtained in Refs~\cite{Bornyakov:1991se} and~\cite{Ejiri:1995gd}.
The behavior of the density at low temperature was not discussed in earlier papers~\cite{D'Alessandro:2007su,D'Alessandro:2010xg,Bornyakov:2011eq}.

The results for a range of temperatures $T_{c}<T\lesssim1.5T_{c}$ obtained on lattices with $L_t=6$ are presented In FIG.~\ref{fig:rho_low_T}.
It can be seen that the thermal monopole density monotonously increases as temperature grows.
The density at $T/T_{c}=1.5$ is $2.3$ times as large as one at $T/T_{c}=1.03$ what indicates that the monopole density is considerably sensitive to temperature.
This fact allows us to say that previous results on this observable did not reflect real situation and the conclusion made in Ref.~\cite{Chernodub:2006gu}
was incorrect.~In FIG.~\ref{fig:rho_low_T} we also show the results from Ref.~\cite{D'Alessandro:2010xg} for comparison.
The results demonstrate similar dependence on temperature, however our results are lower at any temperature.
It is due to Gribov copy effect.

In FIG.~\ref{fig:density} we show our data for temperatures $T>1.5T_{c}$.
The data of Refs.~\cite{D'Alessandro:2007su} and ~\cite{Bornyakov:2011eq} are also shown for comparison.
As was concluded in Ref.~\cite{Bornyakov:2011eq} effects of Gribov copies in results of Ref.~\cite{D'Alessandro:2007su} are between 20\% at $T/T_c=2$ and almost $30\%$ for $T/T_c=7$.~Our results deviate from those of Ref.~\cite{Bornyakov:2011eq} by about $10\%$.
Thus we observe violation of universality of the thermal monopole density at given lattice spacing.
We need to check whether the continuum limit is reached in case of the Symanzik action for our lattice spacings.
To answer to this question we compare results obtained on $L_t=4$ and $L_t=6$ lattices at at $T=1.5~T_{c}$ and $T=2~T_{c}$.
One can see from FIG.~\ref{fig:density} that the change of the density with decreasing lattice spacing is small for both temperature values.
Quantitatively it is less than $1\%$ for $T=2~T_{c}$ and about $3\%$ for $T=1.5~T_{c}$.
This allows us to state that in case of Symanzik action the results for the monopole density obtained on $L_t=4$ lattices can be considered as being close to the continuum limit.
Similar conclusion about closeness to the continuum limit was made for the Wilson action in Ref.~\cite{Bornyakov:2011eq}.
Thus the  violation of universality seen in FIG.~\ref{fig:density} may persist to continuum limit.
Note that at zero temperature the universality breaking effects were found to be much stronger~\cite{Bornyakov:2005iy}, up to $30\%$.

Different values for monopole density in case of Symanzik and Wilson actions can be explained by a fact that configurations have different number of short range dipoles at the same temperatures.
By short range dipoles we imply paired monopole trajectories having opposite wrappings and separated by distance of order of one lattice spacing.
Thus, if we calculate monopole density omitting small distance,it should bring densities closer to each other.
We found indeed that the distance between two densities decreases after we remove dipoles of size $a$.

We will come back to discussion of the universality for the thermal monopole density in the next section.

\begin{figure}[tb]
\vspace*{1.5cm}
\centering
\includegraphics[width=6.0cm,angle=270]{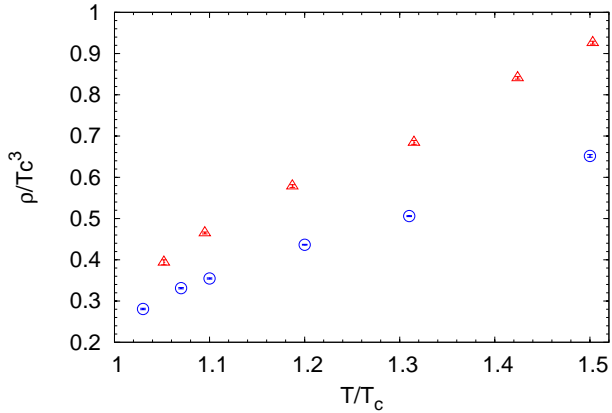}
\caption{The behavior of the monopole density(normalized by $T_{c}^{3}$) at low temperatures(blue empty circles).
~The results from Ref.~\cite{D'Alessandro:2010xg} are presented for comparison(red empty triangles).
}
\label{fig:rho_low_T}
\end{figure}

The dimensional reduction suggests for the density $\rho$ the following temperature dependence at high enough temperature:

\begin{equation}
\rho(T)^{1/3} = c_\rho g^2(T) T
\label{eq:dim reduction}
\end{equation}

\noindent where the temperature dependent running coupling $g^2(T)$ is described at high temperature by the two-loop expression with the scale parameter $\Lambda_T$:

\begin{equation}
g^{-2}(T) = \frac{11}{12\pi^2} \ln (T/\Lambda_T) + \frac{17}{44\pi^2}(\ln(2\ln (T/\Lambda_T))
\label{eq:two_loop_approx}
\end{equation}

We fit our data to function determined by equations (\ref{eq:dim reduction}) and (\ref{eq:two_loop_approx}).
The good fit with $\chi^2/dof = 0.28$ was obtained for $T \ge 2T_c$. The values for fit parameters were $c_\rho=0.160(6)$,~$\Lambda_T/T_c= 0.144(3)$.
Thus for high enough temperature the density $\rho(T)$ is well described by the form which follows from dimensional reduction.
The values of parameters $c_\rho$ and $\Lambda_T$ differ from values obtained in~\cite{Bornyakov:2011eq} though the difference in results for the density is small.
But note that we used fit over range of temperatures between $2T_c$ and $3T_c$ while in~\cite{Bornyakov:2011eq} $T \ge 3T_c$ were used.
The data for the thermal monopole density are presented in TAB.~\ref{tab:rho}.

\begin{figure}[tb]
\vspace*{1.5cm}
\centering
\includegraphics[width=6.0cm,angle=270]{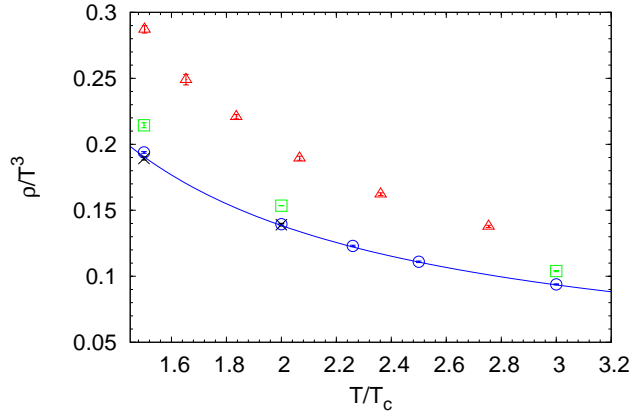}
\caption{The dependence of the thermal monopole density(normalized by $T^{3}$) on temperature(circles).
~The line is a fit to eq.(\ref{eq:two_loop_approx}).~The data from Ref.~\cite{D'Alessandro:2007su}(triangles) and
from Ref.~\cite{Bornyakov:2011eq}(squares) are presented for comparison.~The densities at $1.5T_{c}$ and $2T_{c}$ with $L_{t}=6$ are
labelled by the crosses.
}
\label{fig:density}
\end{figure}

\section{Monopole Interaction}

We computed two types of monopole density correlators $g(r)$, for monopoles having the same charges (MM correlator) and for monopoles having opposite charges (AM correlator).
The correlators are defined as follows:
\begin{equation}
g_{MM}(r) = \frac{\langle\rho_M(0)\rho_M(r)\rangle}{\rho_M^2} + \frac{\langle\rho_A(0)\rho_A(r)\rangle}{\rho_A^2}
\label{eq:mm_corr}
\end{equation}

\begin{equation}
g_{AM}(r) = \frac{\langle\rho_A(0)\rho_M(r)\rangle}{\rho_A\rho_M} + \frac{\langle\rho_M(0)\rho_A(r)\rangle}{\rho_A\rho_M}
\label{eq:am_corr}
\end{equation}

\noindent where $\rho_{M,A}(0)$ and $\rho_{M,A}(r)$ are local densities.
It can be reexpressed as:

\begin{equation}
g_{MM}(r) = \frac{1}{\rho_M}\frac{1}{4\pi r^{2}} \langle\frac{dN_M(r)}{dr}\rangle +\frac{1}{\rho_A}\frac{1}{4\pi r^{2}} \langle\frac{dN_A(r)}{dr}\rangle  \,,
\label{eq:shell_density}
\end{equation}

\noindent where $N_M(r)$ is the number of monopoles, and similarly for $g_{AM}$.

In our computations following Refs.~\cite{D'Alessandro:2007su,Bornyakov:2011eq} we take $dN(r)$ to be a number of monopoles in a spherical shell of finite thickness $dr=a$ at a distance $r$ from a reference particle, whereas $4 \pi r^{2}dr$ is equal to a volume of this shell, i.e. number of lattice sites in it.

Correlators $g_{MM,AM}(r)$ were calculated for nine temperatures in the range between $1.03T_{c}$ and $3T_{c}$.
Three AM and MM correlators are presented in FIG.~\ref{fig:ma_corr} and FIG.~\ref{fig:mm_corr} respectively.
\begin{figure}[tb]
\vspace*{1.5cm}
\centering
\includegraphics[width=6.0cm,angle=270]{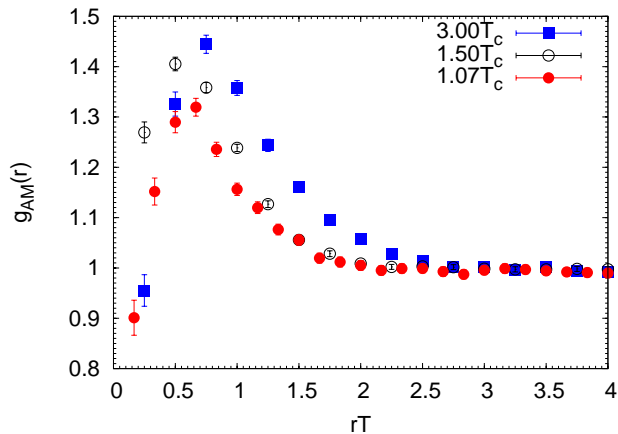}
\caption{
The correlation function $g_{AM}(r)$ for  monopole-antimonopole case for three values of temperature: $3T_{c}$, $1.5�_{c}$ and $1.07T_{c}$ with $L_{t} = 4,4$ and $6$ respectively.
}
\label{fig:ma_corr}
\end{figure}

\begin{figure}[tb]
\vspace*{1.5cm}
\centering
\includegraphics[width=6.0cm,angle=270]{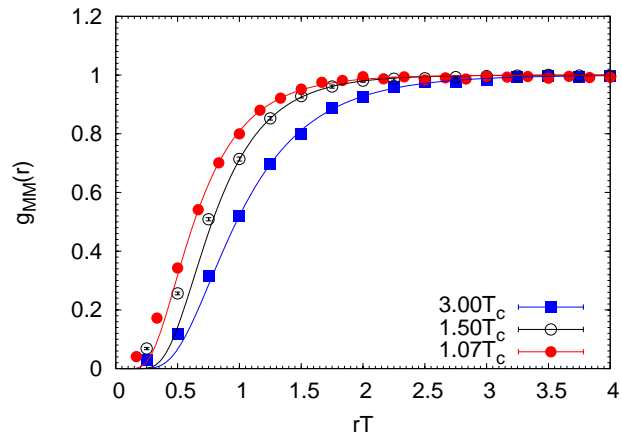}
\caption{
The correlation function $g_{MM}(r)$ for monopole-monopole case for the same values of temperature as in Fig.~\ref{fig:ma_corr}.
The lines are fits for these three cases.
}
\label{fig:mm_corr}
\end{figure}

One can see that in AM case the correlators start below 1, reach  maximal value equal to $1.3 - 1.4$ at distance $r$ between $0.5/T$ and $0.8/T$ and then decrease down to $1$ at distance between $1.8/T$ and $2.5/T$.
In other words the repulsion at short distances changes to attraction at large distances.
In MM case a picture is different, the correlators start at small value below 1 and smoothly reach 1 from below at same distances as AM correlators do.
In this case we can see only the repulsion.
These results are in a good qualitative agreement with results obtained in Refs.~\cite{D'Alessandro:2007su,Bornyakov:2011eq}.

It should be mentioned that we applied more correct procedure to compute MM and AM correlators.
In Refs.~\cite{Bornyakov:2011eq} and~\cite{D'Alessandro:2010xg} only one time slice was used to calculate the correlators since the monopole coordinates could not be determined for all time slices unambiguously because of small loops attached to monopole trajectory.
We used a special algorithm which cut off all UV loops out of each monopoles trajectory.
Then we could take into consideration all time slices to compute the
correlators.
This leads to more precise results for the correlators.
For example, at $T=3T_c$ the statistical errors decreased by factor 1.7 approximately.

One can also see that correlators in FIG.~\ref{fig:ma_corr} and FIG.~\ref{fig:mm_corr} are temperature dependent.
~It is possible to fit the data with the temperature dependent potential:

\begin{equation}
g_{MM,AM}(r)  = e^{-V(r)/T}
\label{eq:corr_fit}
\end{equation}
\noindent where  $V(r)$ can be well approximated by a screened potential:
\begin{equation}
V(r)  = \frac{\alpha_{m}}{r}e^{-m_{D}r}
\label{eq:poten}
\end{equation}
\noindent at large distances. In eq.~(\ref{eq:poten}) $m_D$ is a screening mass and $\alpha_{m}$ is a magnetic coupling.

To decide which data points can be included in the fit range we used the following method.
We plot the dependence of $rV(r)$ on $rT$ in log scale for all temperatures (see FIG.~\ref{fig:log_V(r)}).~The linear dependence
should be valid at distances where  eq.~(\ref{eq:poten}) is satisfied. Thus data at small distances which do not fall onto a line are to be
discarded.~Typically we discarded three data points.~Data at large distances were discarded  because of high statistical errors.
Our results for the  fit parameters are presented in Table~\ref{tab:a_and_m}.~Note that our values for $\chi^2/N_{dof}$  shown in Table~\ref{tab:a_and_m} are in general lower than values of this quantities characterizing fit quality obtained  in  Refs.~\cite{Liao:2008jg,Bornyakov:2011eq}.

\begin{figure}[tb]
\vspace*{1.5cm}
\centering
\includegraphics[width=6.0cm,angle=270]{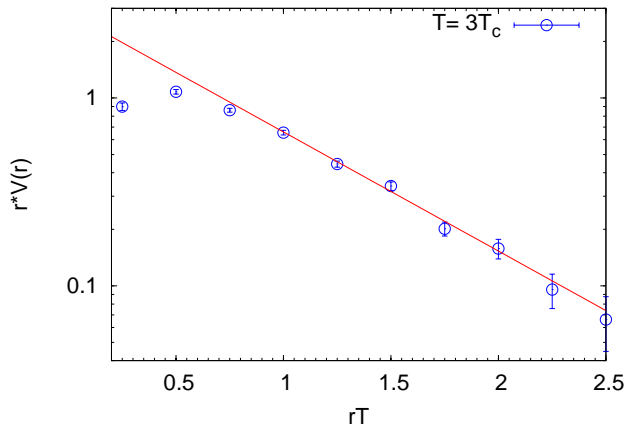}
\caption{The dependence of $rV(r)$ on $rT$.~Note log scale for x axis.~The red line is the fit to eq.~(\ref{eq:poten}).
}
\label{fig:log_V(r)}
\end{figure}

\begin{table}[ht]
\begin{center}
\begin{tabular}{|c|c|c|c|c|c|c|}
\hline
$ T/Tc $&$ \alpha_{m} $&$m_{D}/T $&$\chi^{2}$\\
\hline
\hline
3.00    & 2.84(21)   & 1.47(07)    & 0.81 \\
2.50    & 3.03(31)   & 1.67(08)    & 0.89\\
2.26    & 3.18(24)   & 1.79(06)    & 0.56\\
2.00    & 2.61(15)   & 1.75(05)    & 0.30\\
1.50    & 2.86(15)   & 2.13(05)    & 0.16\\
\hline
\hline
2.00    & 2.19(18)    & 1.43(08)   & 0.82\\
1.50     & 2.47(12)   & 1.87(05)   & 0.16\\
1.20    & 2.04(14)    & 2.06(08)   & 0.73\\
1.10    & 1.43(08)    & 1.81(07)   & 0.44 \\
1.07    & 1.53(07)    & 1.97(05)   & 0.23\\
1.03    & 1.76(10)    & 2.24(7)    & 0.26\\

\hline
\end{tabular}
\end{center}
\caption{Values of the magnetic coupling $\alpha_{m}$ and the screening mass
$m_{D}$ obtained by fitting of MM correlators to eq.~(\ref{eq:corr_fit}).~The double line separates the parameters obtained for $L_{t} = 4$(above the line) from those obtained for $L_{t} = 6$(below the line).
}
\label{tab:a_and_m}
\medskip
\noindent
\end{table}

\begin{figure}[tb]
\vspace*{1.5cm}
\centering
\includegraphics[width=6.0cm,angle=270]{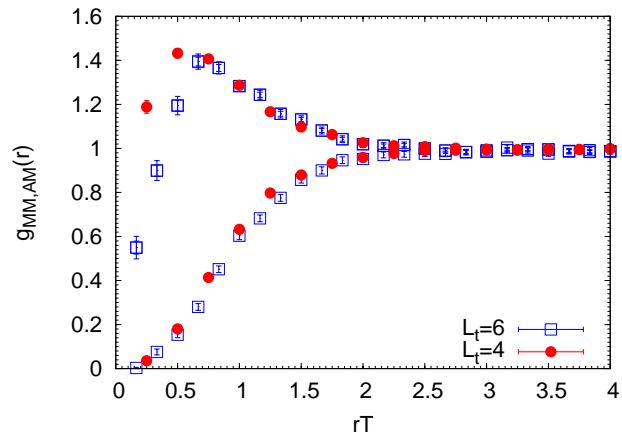}
\caption{Comparison of MM and AM correlators computed at $T=2T_{c}$ on lattices with for $L_{t} = 6$ (empty blue squares) and $L_{t} = 4$ (filled red circles).
}
\label{fig:am_mm_corr_2Tc}
\end{figure}

In order to check finite lattice spacing effects we compared correlators and respective fit parameters on lattices with
$L_{t}=4$ and 6 at $T=1.5T_{c}$ and $2T_c$.
The comparison of these correlators for  $T=2T_{c}$ is presented in FIG.~\ref{fig:am_mm_corr_2Tc}.
~One can see that for $g_{MM}(r) $ the  finite lattice spacing effects are small at all distances being maximal at distances $rT \approx 1.2$.
For $g_{AM}(r) $ the good agreement is also observed at distances to the right of the distance corresponding to the maximum
of the correlator while at short distances the finite lattice spacing effects  are large. Results for $T=1.5T_{c}$ are similar.

\begin{figure}[tb]
\vspace*{1.5cm}
\centering
\includegraphics[width=6.0cm,angle=270]{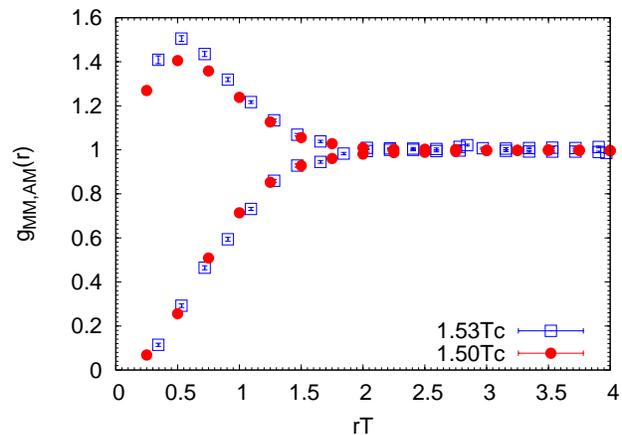}
\caption{
Comparison of MM and AM correlators computed at $T=1.5T_{c}$ in this work(red filled circles) and Ref.~\cite{Bornyakov:2011eq}(blue empty squares).
}
\label{fig:am_mm_corr_vs_Braguta}
\end{figure}

In FIG.~\ref{fig:am_mm_corr_vs_Braguta} we compare the MM and AM correlators computed in this work and in Ref.~\cite{Bornyakov:2011eq} at $T\sim1.5T_c$. It can be seen that the correlators are in good agreement with exception for AM correlator at small distances.~This implies that we observe universality of the correlators apart from distribution of the small dipoles: with Wilson action we observe more such dipoles than with improved action.~Above we have concluded that the number of small dipoles is decreasing with decreasing lattice spacing.~Thus, we may expect that in the continuum limit the universality will restore at small distances as well.

The dependence of the fit parameters, $m_{D}$ and $\alpha_{m}$, on temperature, obtained for MM correlators, is presented in FIG.~\ref{fig:m_param} and FIG~\ref{fig:a_param}, respectively.
One can see that the behavior of the fit parameters at $T$ close to $T_c$ is different from their behavior at high temperature.
Close to $T_c$, in the range between $1.03 T_c$ and $1.1 T_c$ we observe decreasing of both $m_D$ and $\alpha_{m}$.
Such behavior was not observed before.
This indicates that just above the transition the monopole interaction becomes weaker with increasing temperature.
Since near the phase transition the finite volume effects might be strong this
observation should be verified  on larger lattices.
At a bit higher temperature, $T=1.2 T_c$, both parameters jump to higher values and then their dependence on the temperature becomes different.
While for $m_D$ we observe slow decreasing, in agreement with results of Ref.~\cite{Bornyakov:2011eq},magnetic coupling $\alpha_{m}$ is slowly increasing.
This is in qualitative agreement with Refs.~\cite{Liao:2008jg,Bornyakov:2011eq}.
Coming to quantitative comparison we find that our values for $m_D$ are almost within error bars although systematically lower than values
reported in Ref.~\cite{Bornyakov:2011eq}.
The values for $\alpha_{m}$ presented in FIG~\ref{fig:a_param} are substantially lower than the values obtained in Ref.~\cite{Bornyakov:2011eq}.
In the temperature range $T \ge 2T_c$ our results for $\alpha_{m}$ are also lower than values obtained in Ref.~\cite{Liao:2008jg}.
The data in FIG~\ref{fig:a_param} indicate that $\alpha_{m}$ approaches its maximum  value of about $3$ at rather small temperature $T \sim 2.5 T_c$.
But for temperatures close to $T_c$ our values of $\alpha_{m}$ are higher than values presented in~\cite{Liao:2008jg}.

The difference in the behavior of the correlator parameters can be explained at least partially by the fact that in this work we eliminated all small loops out of each wrapping cluster.
This gave us an opportunity to determine the monopole location in a given time slice unambiguously and thus, to use all time slices for correlators computation.
This procedure was used in studies of the thermal monopoles for the first time.

We now can compute the plasma parameter $\Gamma$ which is defined as follows:

\begin{equation}
\Gamma  = \alpha_{m}\left( \frac{4\pi\rho}{3T^3} \right )^{1/3}
 \label{gamma}
\end{equation}

$\Gamma$ is equal to ratio of the system potential energy to its
kinetic energy. If $\Gamma << 1$  the system is a weakly coupled
plasma;~if $\Gamma > 1$, it is a strongly coupled plasma. For
$1\leq \Gamma \leq \Gamma_{c} \sim 100$, the system is in a liquid
state. The dependence of $\Gamma$ on temperature is presented in
FIG.~\ref{fig:gamma}. $\Gamma$ is roughly proportional to $\alpha$
since $\rho^{1/3}/T$ varies slowly with temperature (see
FIG.~\ref{fig:rho_low_T} and ~\ref{fig:density}). Thus at small
temperature we observe in FIG.~\ref{fig:gamma} the nonmonotonic
behavior we saw in FIG~\ref{fig:a_param}. At temperatures above
$1.5T_c$ $\Gamma$ can be well approximated by a constant. We
cannot exclude slight increase or slight decrease of $\Gamma$ for
higher temperatures, though. The independence of $\Gamma$ on
temperature was predicted in Ref.~\cite{Liao:2008jg}. But
quantitatively our result is rather different: in
Ref.~\cite{Liao:2008jg} $\Gamma$ was found substantially higher
and approaching a constant value about $5$ at temperatures above
$4T_c$. Despite this quantitative difference we confirm result of
Ref.~\cite{Liao:2008jg} that the thermal monopoles are in a liquid
state at all temperatures.

\begin{figure}[tb]
\vspace*{1.5cm}
\centering
\includegraphics[width=6.0cm,angle=270]{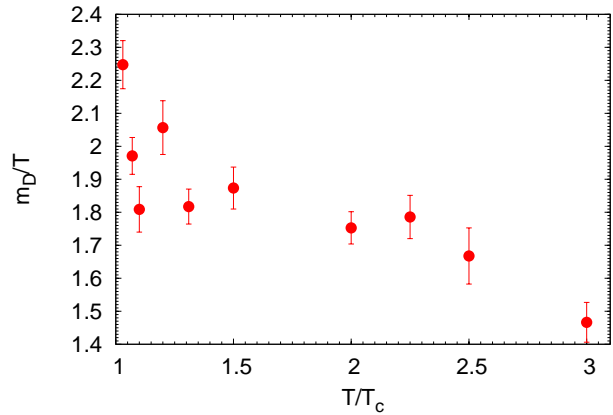}
\caption{
The dependence of the screening mass $m_D$ on temperature.
}
\label{fig:m_param}
\end{figure}

\begin{figure}[tb]
\vspace*{1.5cm}
\centering
\includegraphics[width=6.0cm,angle=270]{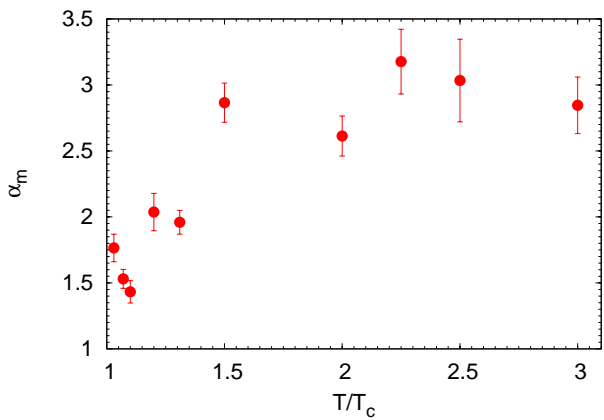}
\caption{
The dependence of the magnetic coupling $\alpha_{m}$ on temperature.
}
\label{fig:a_param}
\end{figure}

\begin{figure}[tb]
\vspace*{1.5cm}
\centering
\includegraphics[width=6.0cm,angle=270]{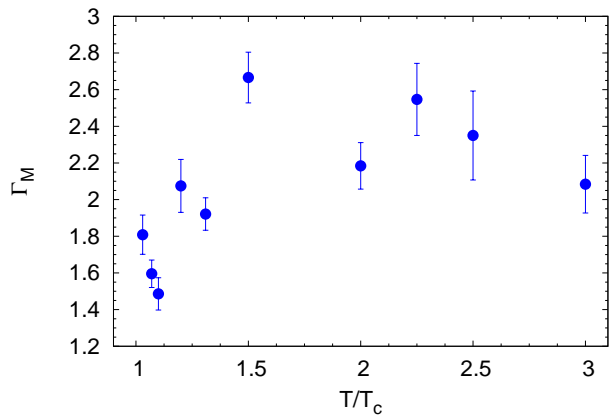}
\caption{
The dependence of the Coulomb plasma parameter on temperature.
}
\label{fig:gamma}
\end{figure}

\section{Monopole Condensation}

In this section we consider thermal monopole trajectories which wrap more than one time in a time direction.
It was proposed in~\cite{D'Alessandro:2010xg} that these trajectories can serve as an indicator of  Bose-Einstein   condensation of the thermal monopoles when the phase transition
is approached from above.~The main idea of this proposal is the following: a trajectory wrapping $k$ times in a time direction is associated with a set of $k$ monopoles permutated cyclically.~Having such an interpretation one can assess a density of these trajectories assuming that they form a system of non-relativistic noninteracting bosons.~According to Ref.~\cite{D'Alessandro:2010xg} this density can be written as follows:

\begin{equation}
\rho_{k}  = \frac{e^{-\hat{\mu} k}}{\lambda^{3}k^{5/2}}
\label{mw_den}
\end{equation}

\noindent where $k$ is a number of wrappings, $\hat{\mu} \equiv -\mu/T$ is a chemical potential, and $\lambda$ is the De Broglie thermal wavelength.~The condensation temperature is determined by the  vanishing of the chemical potential.

To take into account interactions between monopoles it was suggested in~\cite{D'Alessandro:2010xg} to modify eq.~(\ref{mw_den}) to

\begin{equation}
\frac{\rho_{k}}{T^{3}}  = \frac{Ae^{-\hat{\mu} k}}{k^{\alpha}},
\label{mw_den_fit}
\end{equation}

with a free parameter $\alpha$.
The condensation of monopoles still should be signalled by vanishing of effective chemical potential $\hat{\mu}$.

\begin{figure}[tb]
\vspace*{1.5cm}
\centering
\includegraphics[width=6.0cm,angle=270]{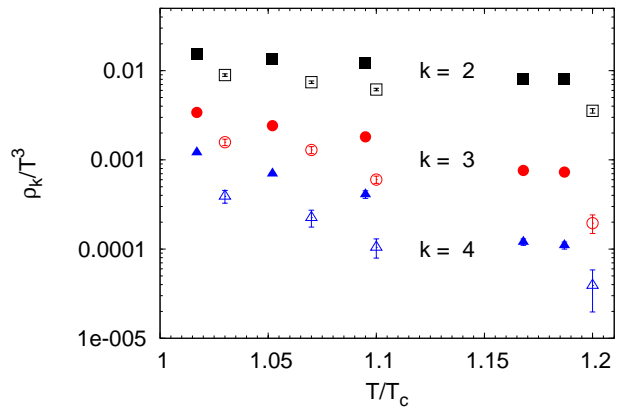}
\caption{Normalized densities for trajectories wrapping more than once in a time direction (empty symbols).~Note a log scale for Y-axes.~For comparison we show here the data from Ref.~\cite{D'Alessandro:2010xg}(filled symbols).
}
\label{fig:mw_density}
\end{figure}

\begin{table}[ht]
\begin{center}
\begin{tabular}{|c|c|c|c|c|}
\hline
$ $                              &$ 1.03Tc$        &$ 1.07Tc       $ &$ 1.1  Tc $      &$ 1.20Tc$   \\
\hline
\hline
$\mu(\alpha = 0)       $         &   1.0(1)        & 1.72(5)         & 2.4(2)          & 2.8(3)    \\
$\mu(\alpha = 2)       $         &   0.50(9)       & 0.99(3)         & 1.6(1)          & 2.0(3)    \\
$\mu(\alpha = 2.5)     $         &   0.37(8)       & 0.82(4)         & 1.4(1)          & 1.8(2)    \\
$\mu(\alpha = 3)       $         &   0.26(7)       & 0.65(5)         & 1.2(1)          & 1.6(2)    \\
\hline
\hline
$\chi^{2}(\alpha = 0)  $         &   2.09          & 0.46           & 2.27           & 1.99        \\
$\chi^{2}(\alpha = 2)  $         &   1.35          & 0.27           & 1.84           & 1.54        \\
$\chi^{2}(\alpha = 2.5)$         &   1.22          & 0.41           & 1.73           & 1.42        \\
$\chi^{2}(\alpha = 3)  $         &   1.14          & 0.64           & 1.61           & 1.30        \\
\hline
\end{tabular}
\end{center}
\caption{Values of fit parameters $\mu$ obtained by fitting data for $\frac{\rho_{k}}{T^{3}}$ to eq.~(\ref{mw_den_fit}) for 4 values of parameter $\alpha$.~Values of $\chi^{2}$ are also shown.
}
\label{tab:mu}
\medskip
\noindent
\end{table}

One can see from FIG.~\ref{fig:mw_density} that our values for $\frac{\rho_{k}}{T^{3}}$ are systematically lower than values presented in Ref.~\cite{D'Alessandro:2010xg}.~This is in a qualitative agreement with the fact reported above that our total density $\rho$ is substantially lower than total density found in  Ref.~\cite{D'Alessandro:2010xg}.
As in the case of the total density this difference is to be explained mostly by large systematic effects due to Gribov copies in results of Ref.~\cite{D'Alessandro:2010xg}.

We fitted our data for $\frac{\rho_{k}}{T^{3}}$ to eq.~(\ref{mw_den_fit}) for temperatures close to $T_c$.
Since our data do not allow us to keep free all parameters we made fits for $4$ fixed values of $\alpha$.
The results of the fit are presented in Table~\ref{tab:mu}.
One can see that $\chi^2$ is decreasing with increasing $\alpha$ for three temperature values out of four.

\begin{figure}[tb]
\vspace*{1.5cm}
\centering
\includegraphics[width=6.0cm,angle=270]{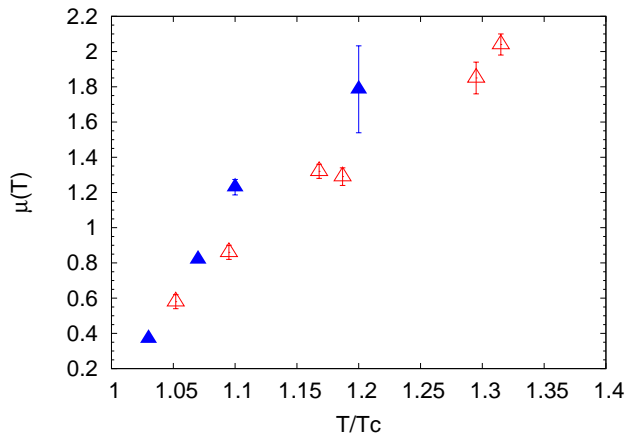}
\caption{The dependence chemical potential on temperature(blue filled triangles).~The data from ref.~\cite{D'Alessandro:2010xg} is presented for comparison(red empty triangles).
}
\label{fig:mu}
\end{figure}

We also computed the monopole thermal mass using the mean squared monopole fluctuation $\Delta r^{2}$ . The relation between $\Delta r^{2}$ and the mass of a nonrelativistic free particle is as follows \cite{D'Alessandro:2010xg}:

\begin{equation}
m = \frac{1}{2T\Delta r^{2}}
\label{m_thermal_def}
\end{equation}

\noindent $\Delta r^{2}$ can be computed on a lattice in the
following way \cite{D'Alessandro:2010xg}:

\begin{equation}
a^{-2}\Delta r^{2} = \frac{1}{L}\sum_{i = 1}^{L}d_{i}^{2}
\label{delta_def}
\end{equation}

\noindent where $L$ - is a total trajectory lengths, $d_{i}^2$ is a squared spatial distance between the monopole coordinate at  $t=0$ and its current coordinate after $i$ steps along the monopole trajectory.

As in Refs.~\cite{D'Alessandro:2010xg,Bornyakov:2011eq} we computed $\Delta r^{2}$ for trajectories with one wrapping.~Furthermore,
we have checked the influence of the small loops on the value of the monopole mass determined via eq.~(\ref{m_thermal_def}).

\begin{figure}[tb]
\vspace*{1.5cm}
\centering
\includegraphics[width=6.0cm,angle=270]{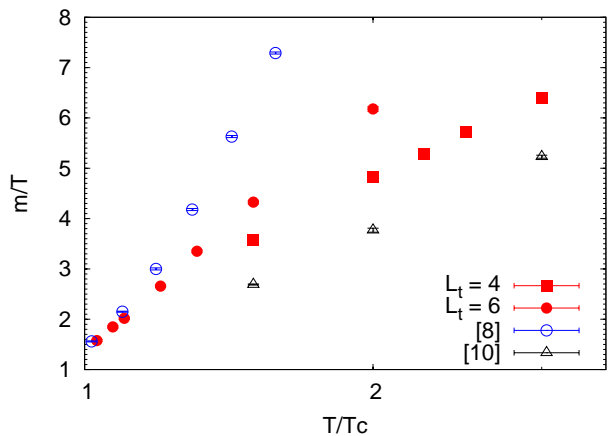}
\caption{
The dimensionless ratio $m/T$ as a function of temperature obtained in our work ($L_{t} = 4$ - red filled squares, $L_{t} = 6$ - red filled circles), in Ref.~\cite{Bornyakov:2011eq}(empty triangles) and in Ref.~\cite{D'Alessandro:2010xg}, $a = 0.047$ (blue empty circles).
}
\label{fig:m_thermal}
\end{figure}

The comparison of the thermal monopole mass obtained for the Wilson action~\cite{D'Alessandro:2010xg,Bornyakov:2011eq} with the values obtained in this paper is presented in FIG.~\ref{fig:m_thermal}
It is seen that the values obtained for the Symanzik action demonstrates the same dependence on temperature as for the Wilson case, but are higher at all temperatures than results of~\cite{Bornyakov:2011eq} and lower than results of~\cite{D'Alessandro:2010xg}.

The monopole mass computed after removal of the loops increases by a factor of $6\%$ at high temperature in comparison with the mass obtained in case when all loops are untouched.
But the difference between two masses is getting larger as temperature approaches to $T_{c}$ reaching $16\%$ at $1.03T_{c}$.
Such behavior of the thermal monopole mass is expectable as with decreasing temperature the number of loops attached to a wrapped cluster increases.
This can be seen from FIG.~\ref{fig:loop_number} where the temperature dependence of number of loops per one wrapped cluster is presented ($r_{lp}$).
When temperature decreases from $3T_{c}$ to $1.03T_{c}$, difference in $r_{lp}$ is one order.

\begin{figure}[tb]
\vspace*{1.5cm}
\centering
\includegraphics[width=6.0cm,angle=270]{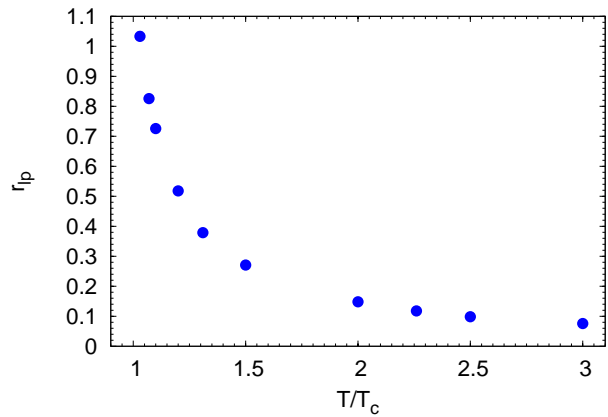}
\caption{
The dependence the average number of loops per one wrapped cluster on temperature.
}
\label{fig:loop_number}
\end{figure}

It was found in~\cite{Bornyakov:2011eq} that the thermal monopole mass can be fitted by the following function:

\begin{equation}
\frac{m}{T} = b\ln(\frac{T}{\Lambda_{m}})
\label{m_thermal_fit}
\end{equation}

The comparison of our results for both cases with the results obtained in Ref.~\cite{Bornyakov:2011eq} are presented in Tab.~\ref{tab:m_thermal_fit_data}.

\begin{table}[ht]
\begin{center}
\begin{tabular}{|c|c|c|c|}
\hline
$b$      & $\Lambda_{m}/T_{c}$ & $ \chi^{2} $ \\
\hline
\hline
3.653(6) & 0.718(2)            & 0.2          \\
3.80(1)  & 0.86(2)             & 0.58         \\
3.66(7)  & 0.78(2)             & 0.26         \\
\hline
\end{tabular}
\end{center}
\caption{
The comparison of the fit parameters for three cases.
The Wilson case (first line)~\cite{Bornyakov:2011eq}, the Symanzik case when loops are untouched(second line) and  the Symanzik case when all loops eliminated(third line).
}
\label{tab:m_thermal_fit_data}
\medskip
\noindent
\end{table}

\section{Conclusions}

We summarize our findings.
Using the improved lattice action (\ref{eq:Symanzik}) and the adequate gauge fixing procedure we completed careful study of the properties of the thermal color-magnetic monopoles.

Comparing our results for thermal monopole density and parameters of the monopole interactions with results of Ref.~\cite{Bornyakov:2011eq}
 we find rather small deviations, at the level of $10\%$ for the density.
We have found that this difference decreases even further when small dipoles are not counted.
This implies universality for infrared thermal monopoles determined in the MAG.
Establishing of the universality of the thermal monopole properties is important to prove that these monopoles
determined after the MAG fixing are fluctuations of the gauge field relevant for infrared physics rather than artifacts of the gauge fixing.

To study cutoff effects we made computations with two lattice spacings (using lattices with $L_t=4$ and 6) at temperatures
$T=1.5T_c$ and $2T_c$.
We find that results for the thermal monopole density are not depending on the lattice spacing (see FIG.~\ref{fig:density}
and Table~\ref{tab:rho}) and thus they are computed in the continuum limit.
Results for parameters of the monopole interaction show slight dependence on the lattice spacing (see Table~\ref{tab:a_and_m}) and
thus they are close to the continuum limit.
The exception is the monopole mass which shows strong dependence on the lattice spacing.

We confirmed observation made before in Ref.~\cite{Bornyakov:2011eq} that without proper gauge fixing the systematic
effects due to Gribov copy effects are large, e.g., up to $30\%$ for the monopole density.
We found that these effects are even more essential for the monopole trajectories with multiple wrappings.

Studying the correlation functions we obtained the values for the Coulomb coupling constant $\alpha_m$ which are substantially smaller than values obtained in Ref.~\cite{Liao:2008jg}.
Since our results are close to the continuum limit and the same is true for results of Ref.~\cite{Liao:2008jg} this difference is to be explained by Gribov copy effects.
Furthermore, we found highly nonmonotonic behavior for both $\alpha_m$ and screening mass $m_D$ near the transition temperature $T_c$.
Both parameters first decrease with increasing temperature and then jump up at the temperature $1.2T_c$ before monotonous dependence on $T$ is settled.
We shall admit that although our measurements in this range of temperature were done on lattices with $L_t=6$, the finite lattice spacing effects should be checked by simulations on lattices with larger value of $L_t$ to check this effect.

Comparatively small values of $\alpha_m$ give rise to small values of the plasma parameter $\Gamma_M$.
We find that this parameter flattens at the value about $2$ at high temperatures in contrast to value of $5$ found in Ref.~\cite{Liao:2008jg}.
Still we confirm that the monopoles are in a liquid state at all temperatures considered.

We have repeated the study of the  monopole trajectories with multiple wrapping undertaken in~\cite{D'Alessandro:2010xg}.
Although we obtained the values for the densities of such trajectories quite different from the values found in~\cite{D'Alessandro:2010xg}, our values for the effective chemical potential $\mu$ are quite close to results of Re.~\cite{D'Alessandro:2010xg}.
Moreover we confirm that $\mu$ goes to zero, indicating Bose-Einstein condensation, at the temperature very close to $T_c$.

We made first studies of the effects of UV fluctuations on parameters of the monopole potential and on monopole thermal mass removing contributions from the closed  loops attached to the wrapped monopole loops.
Due to removing such contribution we were able to identify the currents $j_0$ of the wrapped loop unambiguously and thus to use all time-slices of the lattice in the computation of the correlators, thus decreasing the statistical error substantially.

\subsection*{Acknowledgments}

We would like to express our gratitude to V.V. Braguta, M.I. Polikarpov and V.I. Zakharov for very useful and illuminating discussions.
We also would like to thank both E. D. Merkulova and E. E. Kurshev who helped us a lot with the algorithms used in this work.
This investigation has been supported by the Federal Special-Purpose Programme 'Cadres' of the Russian Ministry of Science
and Education and by grant RFBR 11-02-01227-a.}

\subsection*{Appendix}

In this appendix we present a table of the densities for all studied temperatures.
\begin{center}
\begin{table*}[ht]
\begin{tabular}{|c|c|c|c|c|c|c|c|c|c|}
\hline
$T/Tc$&$\rho_{1}/T^{3}$&$\rho_{2}/T^{3}$&$\rho_{3}/T^{3}$&$\rho_{4}/T^{3}$&$\rho_{5}/T^{3}$&$\rho_{6}/T^{3}$&$\rho_{7}/T^{3}$&$\rho_{8}/T^{3}$&$\rho_{9}/T^{3}$\\
\hline
\hline
$1.03$&$0.245(2)$      &$0.89(3)10^{-2}$&$0.16(1)10^{-2}$&$0.38(6)10^{-3}$&$0.24(5)10^{-3}$&$0.10(3)10^{-3}$&$0.5(2)10^{-4} $&$0.3(2)10^{-4} $&$0.1(1)10^{-4}$\\
$1.07$&$0.261(2)$      &$0.74(2)10^{-2}$&$0.13(1)10^{-2}$&$0.22(5)10^{-3}$&$0.6(2)10^{-4} $&$0.2(1)10^{-4} $&$0.5(3)10^{-4} $&$0.1(1)10^{-4} $&$0.1(1)10^{-4}$\\
$1.10$&$0.260(1)$      &$0.61(2)10^{-2}$&$0.53(6)10^{-3}$&$0.7(2)10^{-4} $&$0.5(2)10^{-4} $&$0.2(1)10^{-4}$ &                &                &               \\
$1.20$&$0.249(1)$      &$0.35(2)10^{-2}$&$0.19(5)10^{-3}$&$0.4(2)10^{-4} $&                &                &                &                &               \\
$1.31$&$0.224(3)$      &$0.21(2)10^{-2}$&$0.4(4)10^{-4}$&                 &                &                &                &                &               \\
$1.50^{a}$&$0.193(6)$  &$0.93(5)10^{-3}$&$0.9(4)10^{-5}$&                 &                &                &                &                &               \\
$1.50^{b}$&$0.1886(12)$&$0.53(7)10^{-3}$&$0.9(9)10^{-5}$&                 &                &                &                &                &              \\
$2.00^{b}$&$0.1390(10)$&$0.11(3)10^{-3}$&               &                 &                &                &                &                &              \\
$2.00^{a}$&$0.139(6)$  &$0.14(2)10^{-3}$&               &                 &                &                &                &                &               \\
$2.26$&$0.123(5)$      &$0.08(2)10^{-3}$&               &                 &                &                &                &                &               \\
$2.5 $&$0.111(1)$      &$0.3(1)10^{-4} $&               &                 &                &                &                &                &               \\
$3.00$&$0.938(4)10^{-1}$&$0.21(6)10^{-4}$&              &                 &                &                &                &                &              \\
\hline
\end{tabular}
\caption{
The monopole density (normalized by $T^{3}$) of monopole trajectories wrapped one and more times in time direction as a function of $T/T_{c}$.
The superscript $a$ and $b$ above the temperature value refers to two different lattice spacing $4$ and $6$ respectively.
}
\label{tab:rho}
\medskip
\noindent
\end{table*}
\end{center}


\begin{thebibliography}{0}
\expandafter\ifx\csname natexlab\endcsname\relax\def\natexlab#1{#1}\fi
\expandafter\ifx\csname bibnamefont\endcsname\relax
  \def\bibnamefont#1{#1}\fi
\expandafter\ifx\csname bibfnamefont\endcsname\relax
  \def\bibfnamefont#1{#1}\fi
\expandafter\ifx\csname citenamefont\endcsname\relax
  \def\citenamefont#1{#1}\fi
\expandafter\ifx\csname url\endcsname\relax
  \def\url#1{\texttt{#1}}\fi
\expandafter\ifx\csname urlprefix\endcsname\relax\def\urlprefix{URL }\fi
\providecommand{\bibinfo}[2]{#2}
\providecommand{\eprint}[2][]{\url{#2}}

\end{thebibliography}


\begin{thebibliography}{28}
\expandafter\ifx\csname natexlab\endcsname\relax\def\natexlab#1{#1}\fi
\expandafter\ifx\csname bibnamefont\endcsname\relax
  \def\bibnamefont#1{#1}\fi
\expandafter\ifx\csname bibfnamefont\endcsname\relax
  \def\bibfnamefont#1{#1}\fi
\expandafter\ifx\csname citenamefont\endcsname\relax
  \def\citenamefont#1{#1}\fi
\expandafter\ifx\csname url\endcsname\relax
  \def\url#1{\texttt{#1}}\fi
\expandafter\ifx\csname urlprefix\endcsname\relax\def\urlprefix{URL }\fi
\providecommand{\bibinfo}[2]{#2}
\providecommand{\eprint}[2][]{\url{#2}}

\bibitem[{\citenamefont{Adams et~al.}(2005)}]{Adams:2005dq}
\bibinfo{author}{\bibfnamefont{J.}~\bibnamefont{Adams}} \bibnamefont{et~al.}
  (\bibinfo{collaboration}{STAR Collaboration}), \bibinfo{journal}{Nucl.Phys.}
  \textbf{\bibinfo{volume}{A757}}, \bibinfo{pages}{102} (\bibinfo{year}{2005}),
  \eprint{nucl-ex/0501009}.

\bibitem[{\citenamefont{Liao and Shuryak}(2007)}]{Liao:2006ry}
\bibinfo{author}{\bibfnamefont{J.}~\bibnamefont{Liao}} \bibnamefont{and}
  \bibinfo{author}{\bibfnamefont{E.}~\bibnamefont{Shuryak}},
  \bibinfo{journal}{Phys. Rev.} \textbf{\bibinfo{volume}{C75}},
  \bibinfo{pages}{054907} (\bibinfo{year}{2007}), \eprint{hep-ph/0611131}.

\bibitem[{\citenamefont{Chernodub and Zakharov}(2007)}]{Chernodub:2006gu}
\bibinfo{author}{\bibfnamefont{M.~N.} \bibnamefont{Chernodub}}
  \bibnamefont{and} \bibinfo{author}{\bibfnamefont{V.~I.}
  \bibnamefont{Zakharov}}, \bibinfo{journal}{Phys. Rev. Lett.}
  \textbf{\bibinfo{volume}{98}}, \bibinfo{pages}{082002}
  (\bibinfo{year}{2007}), \eprint{hep-ph/0611228}.

\bibitem[{\citenamefont{Shuryak}(2009)}]{Shuryak:2008eq}
\bibinfo{author}{\bibfnamefont{E.}~\bibnamefont{Shuryak}},
  \bibinfo{journal}{Prog. Part. Nucl. Phys.} \textbf{\bibinfo{volume}{62}},
  \bibinfo{pages}{48} (\bibinfo{year}{2009}), \eprint{0807.3033}.

\bibitem[{\citenamefont{Ratti and Shuryak}(2009)}]{Ratti:2008jz}
\bibinfo{author}{\bibfnamefont{C.}~\bibnamefont{Ratti}} \bibnamefont{and}
  \bibinfo{author}{\bibfnamefont{E.}~\bibnamefont{Shuryak}},
  \bibinfo{journal}{Phys. Rev.} \textbf{\bibinfo{volume}{D80}},
  \bibinfo{pages}{034004} (\bibinfo{year}{2009}), \eprint{0811.4174}.

\bibitem[{\citenamefont{D'Alessandro and D'Elia}(2008)}]{D'Alessandro:2007su}
\bibinfo{author}{\bibfnamefont{A.}~\bibnamefont{D'Alessandro}}
  \bibnamefont{and} \bibinfo{author}{\bibfnamefont{M.}~\bibnamefont{D'Elia}},
  \bibinfo{journal}{Nucl. Phys.} \textbf{\bibinfo{volume}{B799}},
  \bibinfo{pages}{241} (\bibinfo{year}{2008}), \eprint{0711.1266}.

\bibitem[{\citenamefont{Liao and Shuryak}(2008)}]{Liao:2008jg}
\bibinfo{author}{\bibfnamefont{J.}~\bibnamefont{Liao}} \bibnamefont{and}
  \bibinfo{author}{\bibfnamefont{E.}~\bibnamefont{Shuryak}},
  \bibinfo{journal}{Phys.Rev.Lett.} \textbf{\bibinfo{volume}{101}},
  \bibinfo{pages}{162302} (\bibinfo{year}{2008}), \eprint{0804.0255}.

\bibitem[{\citenamefont{D'Alessandro et~al.}(2010)\citenamefont{D'Alessandro,
  D'Elia, and Shuryak}}]{D'Alessandro:2010xg}
\bibinfo{author}{\bibfnamefont{A.}~\bibnamefont{D'Alessandro}},
  \bibinfo{author}{\bibfnamefont{M.}~\bibnamefont{D'Elia}}, \bibnamefont{and}
  \bibinfo{author}{\bibfnamefont{E.~V.} \bibnamefont{Shuryak}},
  \bibinfo{journal}{Phys. Rev.} \textbf{\bibinfo{volume}{D81}},
  \bibinfo{pages}{094501} (\bibinfo{year}{2010}), \eprint{1002.4161}.

\bibitem[{\citenamefont{Bornyakov and Braguta}(2011)}]{Bornyakov:2011th}
\bibinfo{author}{\bibfnamefont{V.}~\bibnamefont{Bornyakov}} \bibnamefont{and}
  \bibinfo{author}{\bibfnamefont{V.}~\bibnamefont{Braguta}},
  \bibinfo{journal}{Phys.Rev.} \textbf{\bibinfo{volume}{D84}},
  \bibinfo{pages}{074502} (\bibinfo{year}{2011}), \eprint{1104.1063}.

\bibitem[{\citenamefont{Bornyakov and Braguta}(2012)}]{Bornyakov:2011eq}
\bibinfo{author}{\bibfnamefont{V.}~\bibnamefont{Bornyakov}} \bibnamefont{and}
  \bibinfo{author}{\bibfnamefont{V.}~\bibnamefont{Braguta}},
  \bibinfo{journal}{Phys.Rev.} \textbf{\bibinfo{volume}{D85}},
  \bibinfo{pages}{014502} (\bibinfo{year}{2012}), \bibinfo{note}{8 pages, 8
  figures, 3 tables}, \eprint{1110.6308}.

\bibitem[{\citenamefont{'t~Hooft}(1981)}]{'tHooft:1981ht}
\bibinfo{author}{\bibfnamefont{G.}~\bibnamefont{'t~Hooft}},
  \bibinfo{journal}{Nucl.Phys.} \textbf{\bibinfo{volume}{B190}},
  \bibinfo{pages}{455} (\bibinfo{year}{1981}).

\bibitem[{\citenamefont{Kronfeld et~al.}(1987)\citenamefont{Kronfeld, Laursen,
  Schierholz, and Wiese}}]{Kronfeld:1987ri}
\bibinfo{author}{\bibfnamefont{A.~S.} \bibnamefont{Kronfeld}},
  \bibinfo{author}{\bibfnamefont{M.}~\bibnamefont{Laursen}},
  \bibinfo{author}{\bibfnamefont{G.}~\bibnamefont{Schierholz}},
  \bibnamefont{and} \bibinfo{author}{\bibfnamefont{U.}~\bibnamefont{Wiese}},
  \bibinfo{journal}{Phys.Lett.} \textbf{\bibinfo{volume}{B198}},
  \bibinfo{pages}{516} (\bibinfo{year}{1987}).

\bibitem[{\citenamefont{Shiba and Suzuki}(1994)}]{Shiba:1994ab}
\bibinfo{author}{\bibfnamefont{H.}~\bibnamefont{Shiba}} \bibnamefont{and}
  \bibinfo{author}{\bibfnamefont{T.}~\bibnamefont{Suzuki}},
  \bibinfo{journal}{Phys.Lett.} \textbf{\bibinfo{volume}{B333}},
  \bibinfo{pages}{461} (\bibinfo{year}{1994}), \eprint{hep-lat/9404015}.

\bibitem[{\citenamefont{Bornyakov et~al.}(2005)\citenamefont{Bornyakov,
  Ilgenfritz, and Mueller-Preussker}}]{Bornyakov:2005iy}
\bibinfo{author}{\bibfnamefont{V.}~\bibnamefont{Bornyakov}},
  \bibinfo{author}{\bibfnamefont{E.-M.} \bibnamefont{Ilgenfritz}},
  \bibnamefont{and}
  \bibinfo{author}{\bibfnamefont{M.}~\bibnamefont{Mueller-Preussker}},
  \bibinfo{journal}{Phys.Rev.} \textbf{\bibinfo{volume}{D72}},
  \bibinfo{pages}{054511} (\bibinfo{year}{2005}), \eprint{hep-lat/0507021}.

\bibitem[{\citenamefont{Suzuki and Yotsuyanagi}(1990)}]{Suzuki:1989gp}
\bibinfo{author}{\bibfnamefont{T.}~\bibnamefont{Suzuki}} \bibnamefont{and}
  \bibinfo{author}{\bibfnamefont{I.}~\bibnamefont{Yotsuyanagi}},
  \bibinfo{journal}{Phys.Rev.} \textbf{\bibinfo{volume}{D42}},
  \bibinfo{pages}{4257} (\bibinfo{year}{1990}).

\bibitem[{\citenamefont{Woloshyn}(1995)}]{Woloshyn:1994rv}
\bibinfo{author}{\bibfnamefont{R.}~\bibnamefont{Woloshyn}},
  \bibinfo{journal}{Phys.Rev.} \textbf{\bibinfo{volume}{D51}},
  \bibinfo{pages}{6411} (\bibinfo{year}{1995}), \eprint{hep-lat/9503007}.

\bibitem[{\citenamefont{Kitahara et~al.}(1998)\citenamefont{Kitahara, Miyamura,
  Okude, Shoji, and Suzuki}}]{Kitahara:1998sj}
\bibinfo{author}{\bibfnamefont{S.}~\bibnamefont{Kitahara}},
  \bibinfo{author}{\bibfnamefont{O.}~\bibnamefont{Miyamura}},
  \bibinfo{author}{\bibfnamefont{T.}~\bibnamefont{Okude}},
  \bibinfo{author}{\bibfnamefont{F.}~\bibnamefont{Shoji}}, \bibnamefont{and}
  \bibinfo{author}{\bibfnamefont{T.}~\bibnamefont{Suzuki}},
  \bibinfo{journal}{Nucl.Phys.} \textbf{\bibinfo{volume}{B533}},
  \bibinfo{pages}{576} (\bibinfo{year}{1998}), \eprint{hep-lat/9803020}.

\bibitem[{\citenamefont{Ilgenfritz et~al.}(2006)\citenamefont{Ilgenfritz,
  Martemyanov, Muller-Preussker, and Veselov}}]{Ilgenfritz:2006ju}
\bibinfo{author}{\bibfnamefont{E.-M.} \bibnamefont{Ilgenfritz}},
  \bibinfo{author}{\bibfnamefont{B.}~\bibnamefont{Martemyanov}},
  \bibinfo{author}{\bibfnamefont{M.}~\bibnamefont{Muller-Preussker}},
  \bibnamefont{and} \bibinfo{author}{\bibfnamefont{A.}~\bibnamefont{Veselov}},
  \bibinfo{journal}{Phys.Rev.} \textbf{\bibinfo{volume}{D73}},
  \bibinfo{pages}{094509} (\bibinfo{year}{2006}), \eprint{hep-lat/0602002}.

\bibitem[{\citenamefont{Hart and Teper}(1996)}]{Hart:1995wk}
\bibinfo{author}{\bibfnamefont{A.}~\bibnamefont{Hart}} \bibnamefont{and}
  \bibinfo{author}{\bibfnamefont{M.}~\bibnamefont{Teper}},
  \bibinfo{journal}{Phys.Lett.} \textbf{\bibinfo{volume}{B371}},
  \bibinfo{pages}{261} (\bibinfo{year}{1996}), \eprint{hep-lat/9511016}.

\bibitem[{\citenamefont{Bonati et~al.}(2010)\citenamefont{Bonati, Di~Giacomo,
  Lepori, and Pucci}}]{Bonati:2010tz}
\bibinfo{author}{\bibfnamefont{C.}~\bibnamefont{Bonati}},
  \bibinfo{author}{\bibfnamefont{A.}~\bibnamefont{Di~Giacomo}},
  \bibinfo{author}{\bibfnamefont{L.}~\bibnamefont{Lepori}}, \bibnamefont{and}
  \bibinfo{author}{\bibfnamefont{F.}~\bibnamefont{Pucci}},
  \bibinfo{journal}{Phys.Rev.} \textbf{\bibinfo{volume}{D81}},
  \bibinfo{pages}{085022} (\bibinfo{year}{2010}), \eprint{1002.3874}.

\bibitem[{\citenamefont{Arasaki et~al.}(1997)\citenamefont{Arasaki, Ejiri,
  Kitahara, Matsubara, and Suzuki}}]{Arasaki:1996sm}
\bibinfo{author}{\bibfnamefont{N.}~\bibnamefont{Arasaki}},
  \bibinfo{author}{\bibfnamefont{S.}~\bibnamefont{Ejiri}},
  \bibinfo{author}{\bibfnamefont{S.-i.} \bibnamefont{Kitahara}},
  \bibinfo{author}{\bibfnamefont{Y.}~\bibnamefont{Matsubara}},
  \bibnamefont{and} \bibinfo{author}{\bibfnamefont{T.}~\bibnamefont{Suzuki}},
  \bibinfo{journal}{Phys.Lett.} \textbf{\bibinfo{volume}{B395}},
  \bibinfo{pages}{275} (\bibinfo{year}{1997}), \eprint{hep-lat/9608129}.

\bibitem[{\citenamefont{Bornyakov et~al.}(2004)}]{Bornyakov:2003vx}
\bibinfo{author}{\bibfnamefont{V.}~\bibnamefont{Bornyakov}}
  \bibnamefont{et~al.} (\bibinfo{collaboration}{DIK Collaboration}),
  \bibinfo{journal}{Phys.Rev.} \textbf{\bibinfo{volume}{D70}},
  \bibinfo{pages}{074511} (\bibinfo{year}{2004}), \eprint{hep-lat/0310011}.

\bibitem[{\citenamefont{Bali et~al.}(1996)\citenamefont{Bali, Bornyakov,
  M{\"u}ller-Preussker, and Schilling}}]{Bali:1996dm}
\bibinfo{author}{\bibfnamefont{G.~S.} \bibnamefont{Bali}},
  \bibinfo{author}{\bibfnamefont{V.}~\bibnamefont{Bornyakov}},
  \bibinfo{author}{\bibfnamefont{M.}~\bibnamefont{M{\"u}ller-Preussker}},
  \bibnamefont{and}
  \bibinfo{author}{\bibfnamefont{K.}~\bibnamefont{Schilling}},
  \bibinfo{journal}{Phys. Rev.} \textbf{\bibinfo{volume}{D54}},
  \bibinfo{pages}{2863} (\bibinfo{year}{1996}), \eprint{hep-lat/9603012}.

\bibitem[{\citenamefont{Bornyakov et~al.}(1992)\citenamefont{Bornyakov,
  Mitrjushkin, and Muller-Preussker}}]{Bornyakov:1991se}
\bibinfo{author}{\bibfnamefont{V.~G.} \bibnamefont{Bornyakov}},
  \bibinfo{author}{\bibfnamefont{V.~K.} \bibnamefont{Mitrjushkin}},
  \bibnamefont{and}
  \bibinfo{author}{\bibfnamefont{M.}~\bibnamefont{Muller-Preussker}},
  \bibinfo{journal}{Phys. Lett.} \textbf{\bibinfo{volume}{B284}},
  \bibinfo{pages}{99} (\bibinfo{year}{1992}).

\bibitem[{\citenamefont{Ejiri}(1996)}]{Ejiri:1995gd}
\bibinfo{author}{\bibfnamefont{S.}~\bibnamefont{Ejiri}},
  \bibinfo{journal}{Phys. Lett.} \textbf{\bibinfo{volume}{B376}},
  \bibinfo{pages}{163} (\bibinfo{year}{1996}), \eprint{hep-lat/9510027}.

\bibitem[{\citenamefont{Bornyakov et~al.}(2007)\citenamefont{Bornyakov,
  Ilgenfritz, V., M., Mueller-Preussker, and A.I.}}]{Bornyakov:2007}
\bibinfo{author}{\bibfnamefont{V.}~\bibnamefont{Bornyakov}},
  \bibinfo{author}{\bibfnamefont{E.-M.} \bibnamefont{Ilgenfritz}},
  \bibinfo{author}{\bibfnamefont{M.~B.} \bibnamefont{V.}},
  \bibinfo{author}{\bibfnamefont{M.~S.} \bibnamefont{M.}},
  \bibinfo{author}{\bibfnamefont{M.}~\bibnamefont{Mueller-Preussker}},
  \bibnamefont{and} \bibinfo{author}{\bibfnamefont{V.}~\bibnamefont{A.I.}},
  \bibinfo{journal}{Phys.Rev.} \textbf{\bibinfo{volume}{D76}},
  \bibinfo{pages}{054505} (\bibinfo{year}{2007}), \eprint{0706.4206}.

\bibitem[{\citenamefont{Bornyakov et~al.}(2001)\citenamefont{Bornyakov,
  Komarov, and Polikarpov}}]{Bornyakov:2000ig}
\bibinfo{author}{\bibfnamefont{V.}~\bibnamefont{Bornyakov}},
  \bibinfo{author}{\bibfnamefont{D.}~\bibnamefont{Komarov}}, \bibnamefont{and}
  \bibinfo{author}{\bibfnamefont{M.}~\bibnamefont{Polikarpov}},
  \bibinfo{journal}{Phys.Lett.} \textbf{\bibinfo{volume}{B497}},
  \bibinfo{pages}{151} (\bibinfo{year}{2001}), \eprint{hep-lat/0009035}.

\bibitem[{\citenamefont{Bogolubsky et~al.}(2007)\citenamefont{Bogolubsky,
  Bornyakov, Burgio, Ilgenfritz, Mitrjushkin, M{\"u}ller-Preussker, and
  Schemel}}]{Bogolubsky:2007pq}
\bibinfo{author}{\bibfnamefont{I.~L.} \bibnamefont{Bogolubsky}},
  \bibinfo{author}{\bibfnamefont{V.~G.} \bibnamefont{Bornyakov}},
  \bibinfo{author}{\bibfnamefont{G.}~\bibnamefont{Burgio}},
  \bibinfo{author}{\bibfnamefont{E.-M.} \bibnamefont{Ilgenfritz}},
  \bibinfo{author}{\bibfnamefont{V.~K.} \bibnamefont{Mitrjushkin}},
  \bibinfo{author}{\bibfnamefont{M.}~\bibnamefont{M{\"u}ller-Preussker}},
  \bibnamefont{and} \bibinfo{author}{\bibfnamefont{P.}~\bibnamefont{Schemel}},
  \bibinfo{journal}{PoS} \textbf{\bibinfo{volume}{LAT2007}},
  \bibinfo{pages}{318} (\bibinfo{year}{2007}), \eprint{0710.3234}.

\end{thebibliography}

\end{document}